\documentstyle[twocolumn,aps,pre,psfig]{revtex}
\def\figsiz1{5cm}
\def\frontmatter{}

\begin{document}

\def\figurePositionI{
\begin{figure}[thb]
\centerline{\psfig{figure=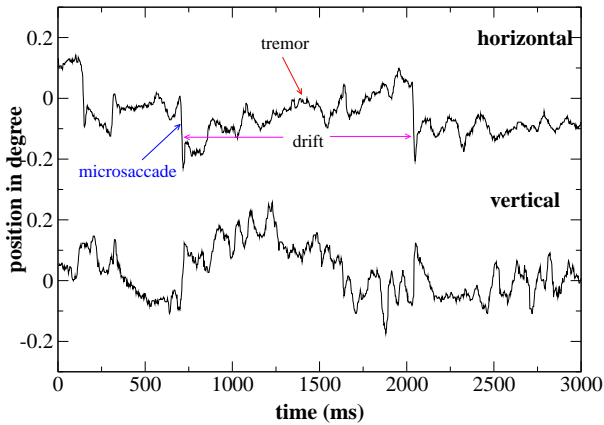,angle=-90,width=9cm}} {
\vspace*{0.0truecm}
 \caption{\label{position86hlvl} \small{Eye
position simultaneous recording of horizontal and vertical
components of left eye
 movements.
  The traces show
 microsaccades, drift and tremor in eye position. In the horizontal tracing, up represents right and
down represents left; in the vertical tracing, up represents up
and down represents down movements.
 }
}}
\end{figure}
}

\def\figureFluctuationI{
\begin{figure}[thb]
\centerline{
\psfig{figure=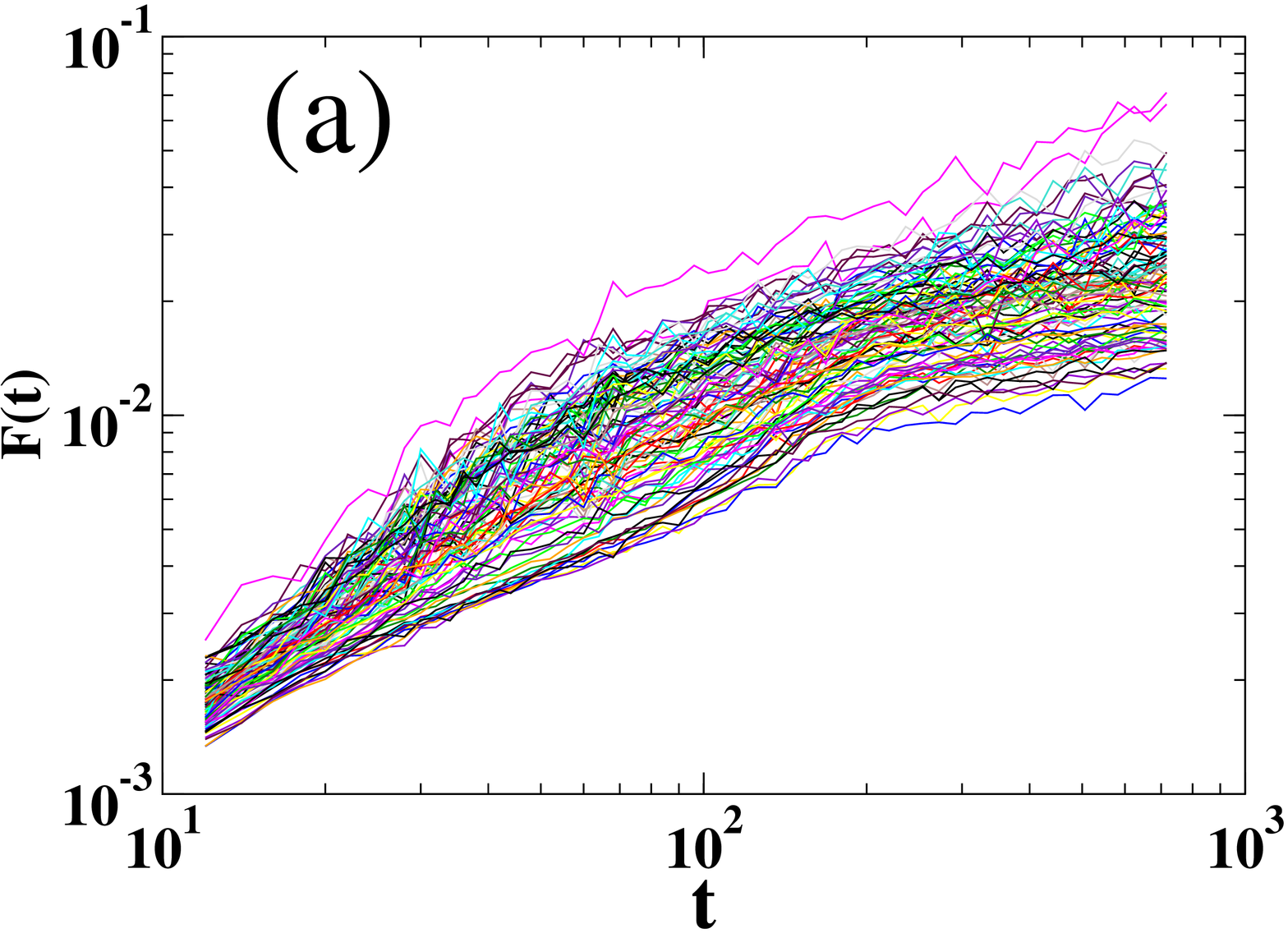,angle=-90,width=4.85cm}
 \psfig{figure=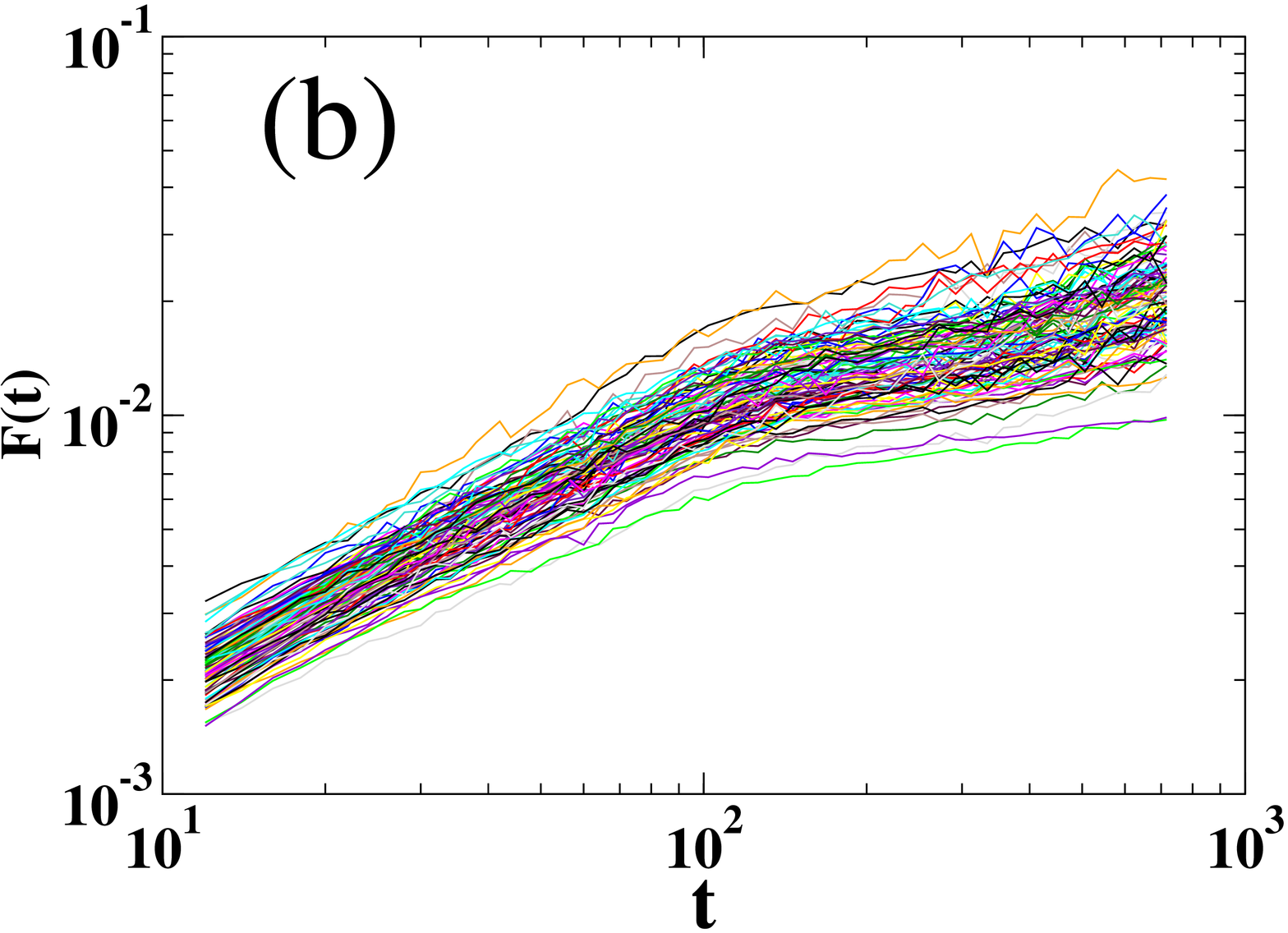,angle=-90,width=4.85cm}
 }
\centerline{
\psfig{figure=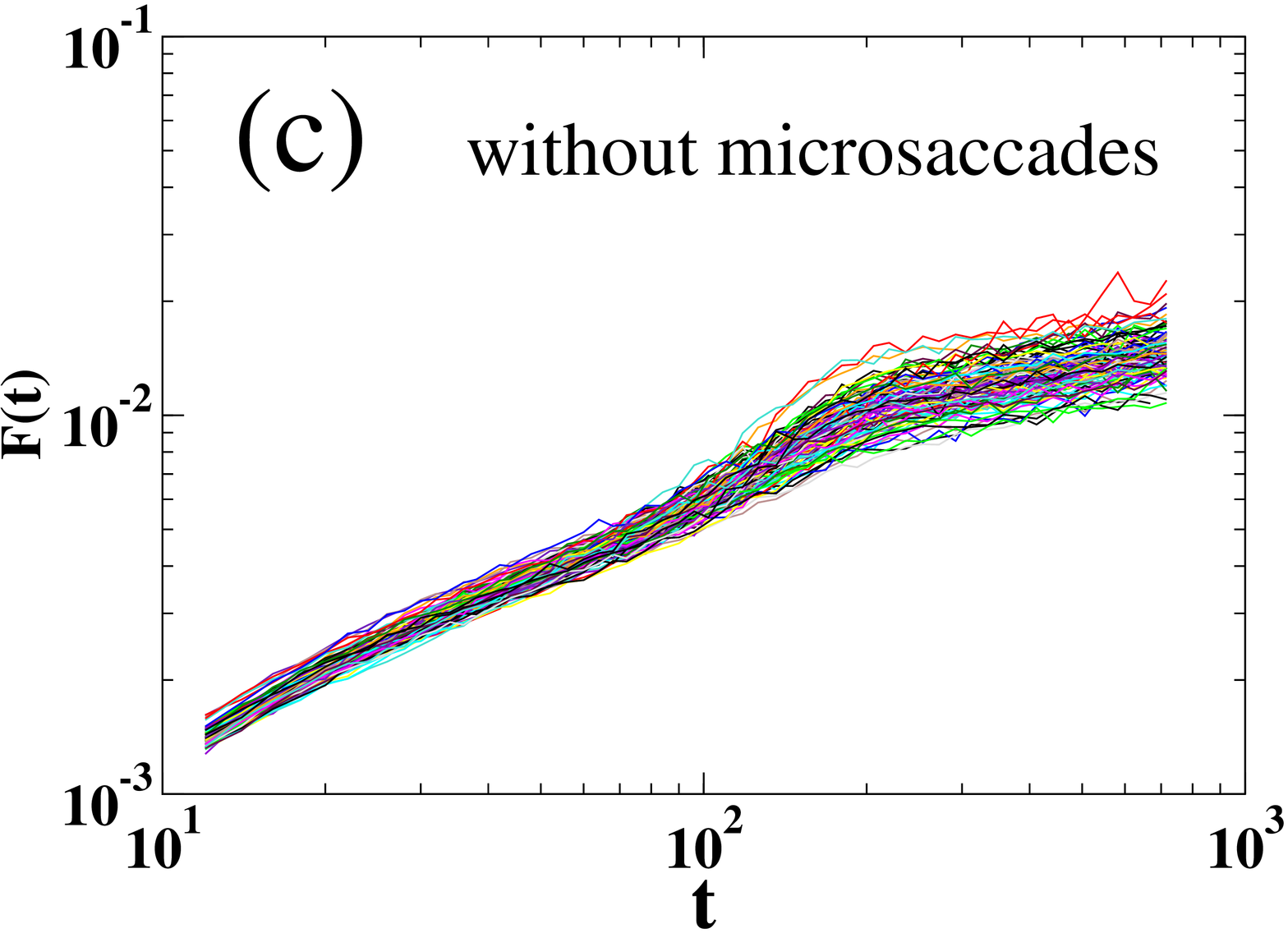,angle=-90,width=4.85cm}
\psfig{figure=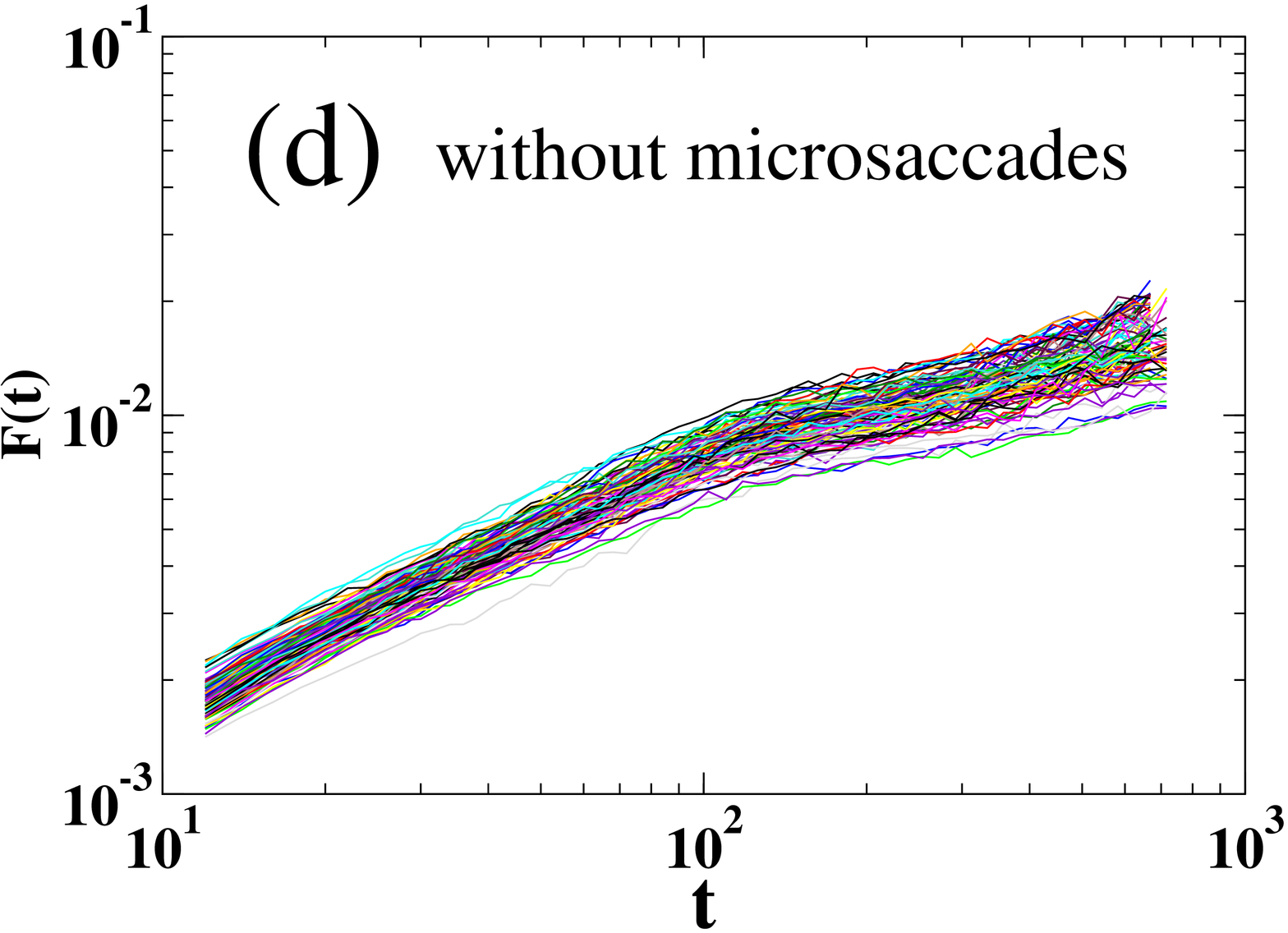,angle=-90,width=4.85cm}
 }
{ \vspace*{0.0truecm} \caption{\label{fluctuationI} \small{
Fluctuation functions obtained by DFA2 for horizontal and vertical
eye movements from the right eye of a typical participant: (a)
horizontal; (b) vertical; (c) horizontal [same data as (a)] after
removing microsaccades; (d) vertical [same data as (b)] after
removing microsaccades. } }}
\end{figure}
}

\def\figureBAfterDFA{
\begin{figure}[thb]
\centerline{\psfig{figure=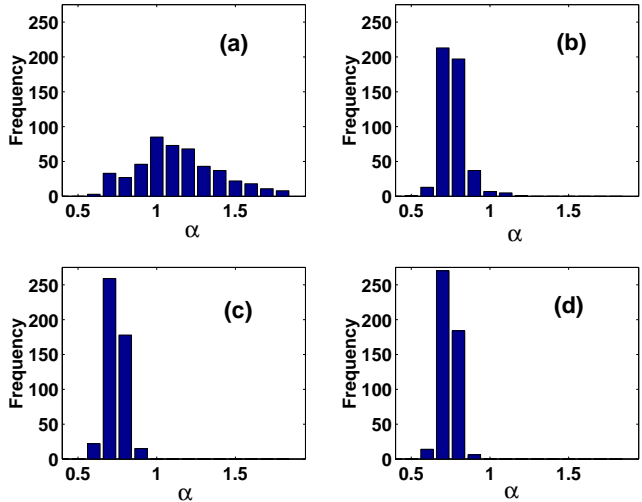,width=8.5cm}}
{ \vspace*{0.0truecm}
 \caption{\label{bafterdfa} \small{Histograms
of the scaling exponents $\alpha$ obtained by DFA2 at the short
time scales [12, 40] ms
 for all the horizontal and
vertical trials (with duration of 3s) with and without
microsaccades
 from  the left eyes of all participants: (a) horizontal; (b) vertical;
 (c) horizontal without
microsaccades; (d) vertical without microsaccades. } }}
\end{figure}
}

 \def\tablepaperDFA{
\begin{table}
\caption{\label{table1} Average values of the scaling exponents
obtained by DFA2 for all fixational eye movements we measured. HL
= horizontal movements of left eyes, HR = horizontal movements of
right eyes, VL = vertical movements of left eyes, and VR =
vertical movements of right eyes. }
\begin{center}
\begin{tabular}{|c|c|cc|}
\multicolumn{2}{|c|}{component}&  Short time scale  &    Long time
scale\\ \hline
&  HL  & 1.13 $\pm$ 0.26  &   0.29 $\pm$ 0.14 \\
A & HR  & 1.05 $\pm$ 0.25  &  0.31 $\pm$ 0.14 \\
& VL  & 0.76 $\pm$ 0.08  &  0.34 $\pm$ 0.13\\
& VR  & 0.76 $\pm$ 0.09  &  0.30 $\pm$ 0.12\\ \hline
\multicolumn{4}{r} {   Microsaccades \ \   removed \quad \quad
\quad \quad}
\\\hline
& HL & 0.74 $\pm$ 0.06 & 0.26$\pm$0.11\\
B &HR  & 0.73 $\pm$ 0.05  & 0.26$\pm$0.10 \\
& VL & 0.74 $\pm$ 0.05 & 0.36$\pm$0.14 \\
& VR & 0.74 $\pm$ 0.04 & 0.35$\pm$0.13 \\
 \end{tabular}
\end{center}
\end{table}
}

\def\figurePowerspec{
\begin{figure}[thb]
\centerline{\psfig{figure=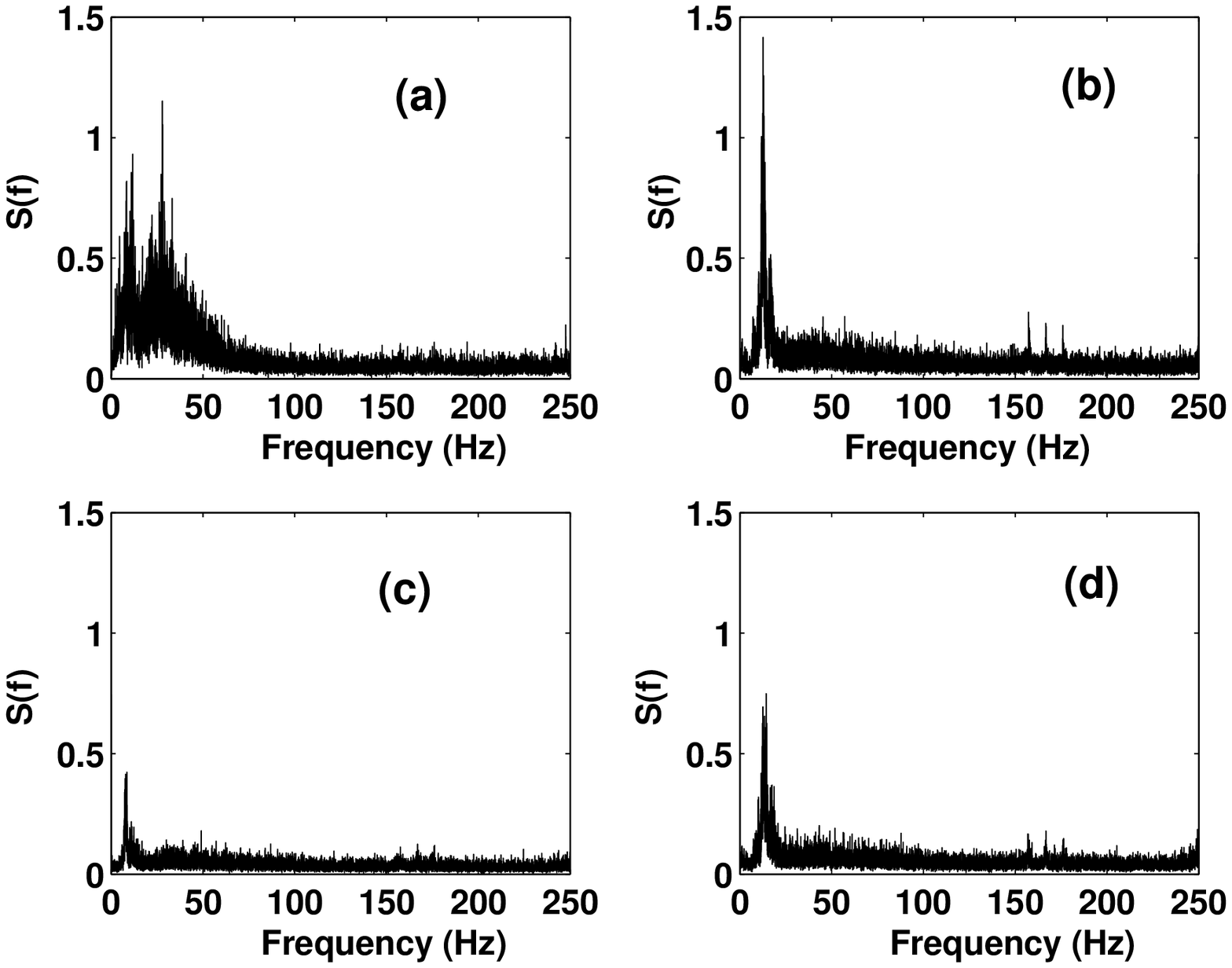,width=8.5cm}}
{
\vspace*{0.0truecm}
\caption{\label{powerspec} \small{Power spectral density for the
velocity series derived from the horizontal and vertical
components from the right eye of one typical participant. (a)
horizontal with microsaccades;
 (b) vertical with microsaccades;
 (c) horizontal after removing the microsaccades; (d) vertical after
 removing the microsaccades
} }}
\end{figure}
}

\title{Scaling of Horizontal and Vertical Fixational Eye
 Movements
 }
\author{Jin-Rong Liang$^{a, b}$
\thanks {jrliang@math.ecnu.edu.cn},
 Shay Moshel $^{a}$,
 Ari Z. Zivotofsky$^{c}$, \\
 Avi Caspi$^{c}$, Ralf Engbert$^{d}$, Reinhold Kliegl$^{d}$ and
 Shlomo Havlin$^{a}$
\thanks {havlin@ophir.ph.biu.ac.il}
  }
\address { $^{a}$Minerva Center and Department of Physics, Bar-Ilan University,
 Ramat-Gan 52900, Israel}
\address { $^{b}$Department of Mathematics, East China Normal University,
  Shanghai 200062, P.R. of China }
\address { $^{c}$Gonda Brain Research Center,  Bar-Ilan University,  Ramat-Gan 52900,
Israel}
\address { $^{d}$Department of Psychology, University of Potsdam, P.O.Box
601553, 14415 Potsdam, Germany
}
\maketitle

\begin{abstract}
 {\bf ABSTRACT \ \

Eye movements during fixation of a stationary target prevent the
adaptation of the photoreceptors to continuous illumination and
inhibit fading of the image. These random, involuntary, small,
movements are restricted at long time scales so as to keep the
target at the center of the field of view. Here we use the
Detrended Fluctuation Analysis (DFA) in order to study the
properties of fixational eye movements at different time scales.
Results show different scaling behavior between horizontal and
vertical movements. When the small ballistics movements, i.e.
micro-saccades, are removed, the scaling exponents in both
directions become similar. Our findings suggest that
micro-saccades enhance the persistence at short time scales mostly
in the horizontal component and much less in the vertical
component. This difference may be due to the need of continuously
moving the eyes in the horizontal plane, in order to match the
stereoscopic image for different viewing distance.

 }
\end{abstract}

\pacs{PACS numbers: }

\frontmatter

\section{Introduction}

When we view a stationary scene, our eyes perform extremely small
autonomic movements. These fixational (or miniature) eye movements
are produced involuntarily and are characterized by three
different types of movements: (i) high-frequency small amplitude
tremor, (ii) slow drift, and (iii) fast microsaccades
\cite{martinez04,moller02}. Generally, they serve to counteract
retinal adaptation by generating small random displacements of the
retinal image in stationary viewing. Studies of fixational eye
movements have been going on since the 1950s, but the role of the
drift, tremor and microsaccadic movements in the visual process is
 not yet fully understood \cite{moller02,engbert03b}.

When fixating an object its image falls on the
fovea, the region of highest visual acuity in the center of the visual field.
 Drifts are slow movements, with a mean amplitude within a range
 of $1.2-9$ min arc  \cite{moller02},
  away from a fixation point.
  Each instance of drift is necessarily terminated by a microsaccade
  (cf. Fig.\ref{position86hlvl}).
 Microsaccades are rapid small amplitude movements ranging between
$1^{\prime}$ and $60^{\prime}$ arc
and occur at a typical mean
rate of 1 to 2 per second \cite{engbert03a}.
 Microsaccades seem to
 reposition the eye on the target.
Tremor (or physiological nystagmus) is a high-frequency (ranging from 50 to
 100 Hz \cite{moller02})
 oscillations of the eye typically less than 0.01 deg, i.e., less than the size
  of one photoreceptor, and is superimposed on drift.
Tremor serves to continuously
 shift the image on the retina, thus calling fresh retinal receptors into
 operation. If an image is artificially fixed
       on the retina it fades and disappears within a few seconds \cite{Riggs}. Tremor causes every
       point of the retinal image to move approximately the distance between
       two adjacent foveal cones in 0.1 seconds and thus
       causes the
image of an object to constantly stimulate new cells in the fovea \cite{olveczky}.
  Drift and tremor movements  are rather irregular  and show statistical
properties of a random walk \cite{engbert03MindEyebook}.
Microsaccades, however, create more linear movement segments
embedded in the eyes' trajectories during fixational movements.
There is evidence that microsaccades are  (i) persistent and
anti-persistent at different time scales \cite{engbert03b},  (ii)
show a characteristic signature of suppression and overshoot in
response to visual change \cite{engbert03a,engbert03c},
 and  (iii) orient themselves according to covert shifts of attention \cite{martinez04}.

Although finding the specific function of fixational
eye movements has been a long-standing and controversial topic in eye
movements research \cite{kowler80}, our concern is not the purpose of such movements but rather
 the dynamical behavior of fixational eye movements
 and if there is some difference between horizontal and vertical fixational eye
  movements.
 In this article we mainly investigate these questions using the
  detrended
  fluctuation
 analysis \cite{peng94,bunde00}, a technique used to detect possible long-term
 correlations in time series.
 We find that
the persistence of horizontal and vertical fixational eye movements
exhibit pronounced different behavior mostly due to the effect of the
  microsaccades. This result is in good agreement with the neurophysiological
fact that horizontal and vertical components of saccades are
controlled by different brain stem nuclei \cite{sparks}. Our study
indicates that after removing the microsaccades the scaling
behavior of both components becomes similar. These findings may
further elucidate the mechanisms underlying effects of
microsaccades on perception and attention
\cite{engbert03b,engbert03a,engbert03c} and their role in the
neurophysiology of vision
  \cite{mart00,mart02,bair98,gres02}.
 In addition, in many pathological states the fixation system can be disrupted
  by slow drift,
nystagmus, or involuntary saccades. However, because all three of these occur
 in healthy individuals it may be difficult to
determine if there is truly an abnormality present. Thus, further
characterizing of the
 fixational system may be useful in clinical
evaluation of such dysfunction.

 \section{DATA}

Data was collected from five normal subjects.
 Eye movements for these participants
were recorded using an EyeLink-II system with a sampling rate of
500 Hz and an instrument spatial resolution $<0.005^{\circ}$. The
subjects were required to fixate a small stimulus with a spatial
extent of $0.12^\circ$ or $7.2$ arc min ($3\times 3$ pixels on a
computer display, black square on a white background). Each
participant performed about 100 trials with a duration of 3
seconds and total of 474 trials were obtained \cite{engbert03b}.
The recording of each trial includes position trajectories of
horizontal and
 vertical components of left eye and right eye movements.

\figurePositionI
 Figure \ref{position86hlvl} shows a typical
simultaneous recording of horizontal and vertical miniature eye
movements for the left eye from one subject. The horizontal and
vertical movements (upper and lower traces in the figure,
respectively) exhibit an alternating sequence of slow drift and
resetting microsaccades. Usually, the subjects show an individual
preponderance regarding the direction of these drifts and
resetting microsaccades. In this subject, for example, the drift
in the horizontal movement occurred typically to the right and the microsaccades to the left
(Fig.\ref{position86hlvl}).

\section{Methods of Analysis}

To study the dynamical behavior of fixational eye movements
we employ the detrended fluctuation analysis (DFA) which was developed
 to quantify long-term power-law correlations embedded in a nonstationary time
series \cite{peng94}. The DFA method has been successfully applied to research
fields such as cardiac dynamics \cite{peng95,bunde00,iva96,iva99,ash01},
human gait \cite{haus95}, climate
temperature fluctuations \cite{bunde98,bunde03} and neural receptors
 in biological systems\cite{bahar01}.
Here we apply this method to the velocity series derived from the position series of
fixational eye movements.
For a position series $x_i $, $ i=1, \cdots, N+1$, of a horizontal or vertical movement,
 we first calculate its velocity series $ v_i$ by
$ v_i = T_0 (x_{i+1}-x_i)$,
 where $T_0$ is the sampling rate; in our experiments
$T_0=500$ Hz. We chose to use a two-point velocity in order to avoid any smoothing 
and  clearly characterize the direction and magnitude
of a movement.  For other definitions of velocity see
\cite{engbert03a}.

 We first calculate the integrated series
  as a profile
\begin{equation}\label{3.1}
Y(k)=\sum_{i=1}^k [v_i-\langle v \rangle],  \  k=1, \cdots, N.
\end{equation}
   Subtraction of the mean $\langle v \rangle$  of the
whole series is not compulsory since it would be eliminated by the
detrending in the third step \cite{kan01}. Thus $Y(k)$ in Eq.(\ref{3.1})
 represents actually
the ``position''.

We then divide the profile $ Y(k) $ of $N$ elements into $N_t =
\mbox{int}(N/t)$ non-overlapping segments of equal length $t$,
where int$(N/t)$ denotes the maximal integer not larger than
$N/t$. Since the length $N$ of the series is often not a multiple
of the considered time scale $t$, a short part at the end of the
profile may remain. In order not
 to disregard this part of the series, the same procedure is repeated starting from
the opposite end. Therefore, $2N_t$ segments  are obtained altogether.

Next, we
determine in each segment the best polynomial fit of the profile and calculate
the variance of the profile from these best polynomials
\begin{equation}\label{3.2}
F^2(\nu,t) \equiv \frac{1}{t} \sum_{i=1}^t \{Y((\nu-1) t+i) -y_\nu(i)\}^2
\end{equation}
for each segment $\nu, \nu=1, \cdots, N_t$, and
\begin{equation}\label{3.3}
F^2(\nu,t) \equiv \frac{1}{t} \sum_{i=1}^t \{Y(N-(\nu-N_t)t+i) -y_\nu(i)\}^2
\end{equation}
for
$ \nu=N_t+1, \cdots, 2N_t$, where $y_\nu$ is the fitting polynomial in
segment $\nu$. If this fitting polynomial is linear, then it is
 the first-order detrended fluctuation analysis (DFA1).
This eliminates the influence of possible linear trends in the profile
on scales larger
than the segment \cite{peng94}.
 In general, in the $n$th order $DFA$ (DFA$n$),
 $y_\nu$ is the best $n$th-order polynomial fit of the profile
 in segment $\nu$. Therefore, linear, quadratic, cubic, or higher order
  polynomials can be
 used in the fitting procedure. Since the detrending of the original time series is done
 by the subtraction of the polynomial fits from the profile, different order DFA
 differ in their capability of eliminating trends of order $n-1$ in the series.

Finally, the fluctuation $F(t)$ over the time windows of size $t$ is determined as
a root-mean-square of  the variance
$$
F(t)=\sqrt{\frac{1}{2N_t} \sum_{\nu =1}^{2N_t} F^2(\nu,t) }.
$$
This computation is repeated over all possible interval lengths.
Of course, in DFA $F(t)$ depends on the DFA order $n$. By
construction, $F(t)$ is only defined for $t \geq n+2.$ For very
large scales, for example, for $t > N/4$, $F(t)$ becomes
statistically
 unreliable because
the number of segments $N_t$ for the averaging procedure becomes
very small. We therefore limit our results to $[n, N/4]$.

  Typically, $F(t)$ increases with interval length $t$.
 We determine the scaling behavior of the fluctuations by analyzing log-log
 plots of $F(t)$ versus $t$.
  A power law $F(t) \propto t^\alpha,$ where $\alpha$ is a scaling
exponent, represents the long-range power-law correlation
properties of the signal. If $\alpha = 0.5$, the series is
uncorrelated (white noise); if $\alpha < 0.5$, the series is
anti-correlated; if $\alpha > 0.5$, the series is correlated or
persistent.

\section{Analysis of Fixational Eye Movements }

We applied DFA1-4 to all velocity records derived from the
horizontal and vertical components. Since the scaling exponents of
the fluctuation functions obtained by DFA1-4 are similar, we show
here the DFA2 results as a representative of the DFA analysis.

 As can be seen from Figs.\ref{fluctuationI} (a) and (b),
the fluctuation functions of horizontal components have pronounced differences
 from the fluctuation function of vertical components.
This is expressed by several characteristics, which can be
observed. There is a broader range of exponents in the horizontal
compared to the vertical. The crossover times, from large
exponents (at short time scales) to smaller exponents (at large
time scales) in horizontal, also show a broader range compared
with the vertical. Moreover, the scaling exponents at short time
scales (between 12 millisecond and 40 milliseconds) for
horizontal, are typical larger than the corresponding exponents
for vertical. However, if we remove microsaccades \cite{FN} the
fluctuations of both components, the corresponding crossovers and
the exponents at small time scales become similar (Figs.
\ref{fluctuationI}(c) and (d)). This result indicates that
microsaccades strongly influence the horizontal components in
fixational eye movements. Note, the close similarity of the
fluctuation function $F(t)$ in the different 3 seconds trials, in
particular after removing the microsaccades, indicates that the
scaling exponent is a stationary and significant characteristic of
the eye movement. \figureFluctuationI

In Fig.\ref{bafterdfa} we show the histograms of the scaling
exponents for the short time scales,
 for all trials with and without microsaccades
 from  the left eyes of all participants.
From this plot we notice that,
 at the short time scale, the horizontal and vertical components
 exhibit persistent behavior  ($\alpha>0.5$) where  the horizontal
components are much stronger correlated than the vertical.
 The average value of the scaling exponents
 for all trials
is $0.76$ for the vertical components and $1.1$ for the horizontal
components (See Table \ref{table1}(A)). The scaling exponents of
horizontal components
 show a broader distribution than the vertical components.
 However, after removing microsaccades,
the fluctuations of horizontal components and the corresponding
scaling exponents have a pronounced change to a narrow
distribution while the vertical components change very little
(see Figs.\ref{fluctuationI} (b) and (d); and Figs.\ref{bafterdfa}
 (b) and (d)).
 When comparing the scaling exponents for the original horizontal
series with the scaling exponents for the horizontal
 removed microsaccades series,
we find that the scaling exponents decrease from an average value
around 1.1 to 0.74, while for the vertical components the scaling
exponents decrease from an average value around 0.76 to 0.74
(Table \ref{table1}). Horizontal and vertical become similar
after removing microsaccades.

 \figureBAfterDFA

We find that at long time scales (between 300 and 600
milliseconds), the horizontal and vertical components show
anti-persistence behaviour ($\alpha<0.5$) (see Table \ref{table1}
(A)), with no significant differences between them.
 After removing the microsaccades  the scaling exponents
at the long time scales, remain almost the same as before. The
horizontal components become slightly less anti-persistent than
vertical (see Table \ref{table1}(A)).

\tablepaperDFA
 We thus conclude that,
 microsaccades in the horizontal components
 are more dominant than in the vertical direction in fixational eye movements.
The microsaccades enhance the persistence mostly in the horizontal
components
 at the short time scales.
At the long time scales both horizontal and vertical components are anti-persistence and
less affected by the microsaccades.

 To further test if the above results are indeed affected by microsaccades,
we  randomly removed parts of the series under study with the same
length as the removed microsaccades and repeated the DFA analysis.
 We found that
this procedure does not influence the scaling exponents.
 Thus, the scaling difference between
the series with and without microsaccades
 is indeed due to microsaccades.

Finally, we tested if the effect of microsaccades can be seen also
in the power spectral density. To this end we analyzed the power
spectra of the horizontal and vertical velocity series, for the
right eye of a typical participant, for all trials with and
without microsaccades. Results are shown in Figs.\ref{powerspec}
(a) and (b) where the microsaccades are included. The
 power spectral density of horizontal and vertical components
 are found to be different (Figs.\ref{powerspec} (a) and (b)).
 After removing the microsaccades the components become similar
(Figs.\ref{powerspec} (c) and (d)). This finding also indicates
that the effect of
 the microsaccades in the horizontal component is stronger than in the vertical.
Note, that this effect is seen much clearer in the DFA curves
where only a few trials (of 3 sec) are sufficient to
distinguish between the horizontal and vertical eye movements.
\figurePowerspec

\section{Discussion}

When the visual world is stabilized on the retina, visual
perception fades as a consequence of neural adaptation. During
normal vision we continuously move our eyes involuntarily even as
we try to fixate our gaze on a small stimulus, preventing retinal
stabilization and the associated fading of vision
\cite{martinez04}. The nature of the neural activity correlated
with microsaccades at different levels in the visual system has
been a long standing controversy in eye-movements research.
Steinman \cite{stei73} showed
 that a person may select
not to make microsaccades, and still be able to see the object of interest,
whereas Gerrits \& Vendrik \cite{gerr70} and Clowes \cite{clow62}
found that optimal viewing
conditions were only obtained when both microsaccades and drifts were
present. Since microsaccades can be suppressed voluntarily in high acuity
observation tasks \cite{bridgeman80,winterson76}, it was concluded that microsaccades serve no useful
purpose and even that they represent an evolutionary puzzle
 \cite{engbert03a,kowler80}.

Our study using DFA suggests that microsaccades play different
roles on different time scales in vertical and horizontal
components in the correction of eye movements, consistent with
\cite{engbert03b}. Moreover we show that due to microsaccades
there is also different scaling behaviour in horizontal and
vertical fixational eye movements. Our results suggest that
microsaccades at short time scales, enhance the persistence mostly
in horizontal movements and much less in the vertical movements.

Our findings that the persistence in horizontal and vertical
fixational eye movements, which are controlled by different brain
stem nuclei, exhibit pronounced different behavior also show that
the role of
 microsaccades in horizontal movements are more dominant.
These findings
may provide  better understanding of the
  recent neurophysiological findings on the effects of
microsaccades on visual information processing
\cite{mart00,mart02,bair98,gres02}.

\vspace*{0.5cm}

We thank Tomer Kalisky for his assistance in applying the DFA
method, and Samuel Ron and Lance M. Optican for constructive
suggestions on the manuscript. This work is supported in part by
grants KL955/6 and KL955/9 (Deutsche Forschungsgemeinschaft) and
by NSFC (No. 10271031) and
 the Shanghai Priority Academic Discipline.


\begin{thebibliography}{00}
\bibitem{FN}
Microsaccades were detected from the velocity series. We remove
velocities above a threshold of 15 deg/sec. On average between 6
and 7 microsaccades are removed for each 3 second trial.
\end{thebibliography}
\end{document}